# MetaHQ: Harmonized, high-quality metadata annotations of public omics samples and studies

Parker Hicks[1], Lydia E Valtadoros[1], Christopher A Mancuso[2], Faisal Alquadoomi[1], Kayla A Johnson[3,4], Sneha Sundar[4], and Arjun Krishnan[1*]

[1]Department of Biomedical Informatics, University of Colorado Anschutz, Aurora, CO 80045; [2]Department of Biostatistics and Informatics, University of Colorado Anschutz, Aurora, CO 80045; [3]Department of Biochemistry and Molecular Biology, Michigan State University, East Lansing, MI 48824; [4]Department of Computational Mathematics, Science and Engineering, Michigan State University, East Lansing, MI 48824

## Abstract

Public omics databases like the Gene Expression Omnibus and the Sequence Read Archive offer substantial opportunities for data reuse to address novel biomedical questions. However, it is still difficult to find samples and studies of interest since they are described by free-text metadata and lack standardized annotations. To address this issue, multiple research groups have undertaken curation efforts to add standardized annotations to large collections of these data, but these annotations are fragmented across online resources and are stored in different formats subject to varying standardization criteria, hindering the integration of annotations across sources. We developed MetaHQ to harmonize and distribute standardized metadata for public omics samples. MetaHQ comprises a database with nearly 200,000 annotations from 13 sources and a user-friendly command-line interface (CLI) to query the database and retrieve annotations. The MetaHQ CLI is deployed as a Python Package on PyPI at https://pypi.org/project/metahq-cli that accesses the MetaHQ database available at https://doi.org/10.5281/zenodo.18462463. Project source code and documentation are available at https://github.com/krishnanlab/meta-hq.

Correspondence: *arjun.krishnan@cuanschutz.edu*

## Introduction

The millions of publicly available transcriptomic profiles across repositories like the Gene Expression Omnibus (GEO) [1] and the Sequence Read Archive (SRA) [2] represent an invaluable resource for transcriptomic data reuse to generate new hypotheses and discoveries [3]. However, effectively finding the appropriate data is severely impeded by the poor quality of metadata that describe samples and studies [4, 5]. Most metadata exist as unstructured plain text where even basic characteristics such as tissue, disease, sex, and age are recorded using inconsistent, non-standardized terminologies that vary dramatically across studies.

To address this challenge, multiple research groups and consortia have invested substantial effort in curating high-quality sample and study annotations using controlled vocabularies and ontologies such as Cell Ontology [6], UBERON [7], and MONDO [8]. However, these curated annotations remain scattered across disconnected online resources: searchable databases [9–11], static project websites [12], GitHub repositories [13], data repositories (e.g., Zenodo, Figshare) [14], and publication supplementary files [15–20]. Moreover, different efforts annotate different attributes with varying quality standards and often exclude connections to important related metadata fields (e.g., series, platform IDs, cross-references). This fragmentation prevents researchers from seamlessly leveraging annotations from multiple curation efforts to retrieve subsets of reliably-annotated data relevant to a specific biological context of interest, with two use cases being to either create large labelled training sets for machine learning (ML) and AI models that can predict sample attributes from molecular profiles [14, 19, 20] or infer standardized annotations from unstructured metadata text [21–23]. Further, for these tasks, in addition to associating annotations to biomedical ontologies, it is also critical to be able to comprehensively and accurately recover samples and studies based on logical, implied annotations. For example, "myocardial cell" samples should be retrieved when querying "heart" and should be labelled as 'NOT' (or 'negative') for unrelated tissues like "liver". Currently, creating such hierarchically propagated annotations or ML-ready labeled datasets is not straightforward and requires domain expertise in biological knowledge graphs. Here we present MetaHQ, a platform that harmonizes and distributes high-quality metadata annotations for over 188,000 publicly available transcriptomics samples and 11,000 studies in GEO. MetaHQ integrates annotations from 13 isolated sources covering tissue, disease, sex, age, and sample type, organized under a unified schema that tracks annotation source and quality. A command-line interface (CLI) enables programmatic querying, ontology-aware retrieval with hierarchical propagation, and generation of ML-ready labeled datasets. By bringing scattered community curation efforts under one roof, MetaHQ transforms researchers' abilities to discover, integrate, and effectively reuse public omics data. Though MetaHQ currently houses harmonized annotations for transcriptomics samples and studies, its general design enables easy expansion to other omics types through additional submissions from the community.

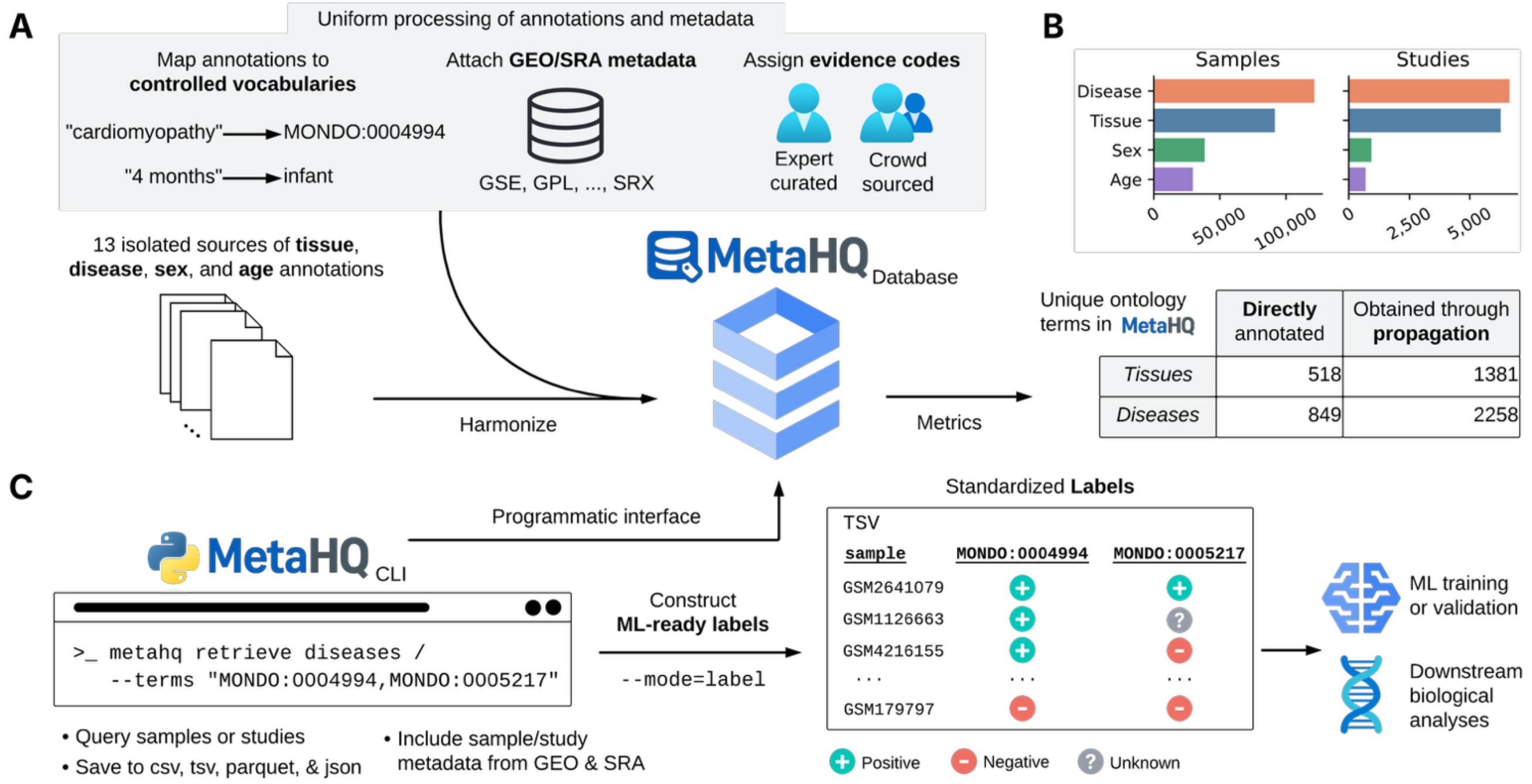

**Figure 1.** Overview of the end-to-end MetaHQ workflow. (A) The harmonization pipeline to combine independent annotation sources into the MetaHQ database. (B) Number of annotations for each attribute type and counts of unique tissue and disease values in the MetaHQ database. (C) The pipeline to programmatically access the database to retrieve labels for downstream applications. Figure created in Lucidchart (www.lucidchart.com).

## Database generation

The MetaHQ database contains annotations from Gemma [10], ALE [13], Bgee [11], CellO [14], CREEDS [12], DiSignAtlas [9], URSA [19], URSA-HD [20], independent curations from Gu *et al.* [17], Johnson and Krishnan [15], Golightly *et al.* [16], Sirota *et al.* [18], and an unpublished in-house curation effort. Together, these sources provide extensive standardized annotations spanning tissue, disease, sex, and age attributes for public transcriptomes. While individual sources annotate only one or two attribute types (Figures S1 and S2), their harmonization enables comprehensive annotation of 188,223 samples across 11,718 studies, covering 1,381 tissues, 2,258 diseases, two sexes, and seven age groups (Figure 1B). While tissue and disease annotations are abundant, annotation coverage is sparse across attributes, with only a small fraction of samples and studies annotated to all four attribute types (Figures S3 and S4).

Lastly, while the database is predominantly human-focused, it also includes curated annotations for mouse, rat, and zebrafish transcriptomes. To our knowledge, MetaHQ is the largest source of harmonized high-quality annotations of public transcriptomes which can be easily queried through a user-friendly CLI (Figure 1C).

## Programmatic interface

### Querying MetaHQ

Users can query MetaHQ annotations for tissue, disease, sex, and age attributes of interest with `metahq retrieve`. They can choose between sample-level queries (`--level=sample`) and study-level queries (`--level=series`, as studies are termed "series" in GEO), and apply additional filters by species, technology (e.g., RNA-Seq or microarray), and evidence code (e.g., expert or crowd). Users can also append standard metadata fields to their query such as platform, description, and cross-references to SRA. Below is an example of a query for disease sample annotations:

```
metahq retrieve diseases --level sample --terms "MONDO:0004994,MONDO:0018177" \
  --filters "species=human,tech=rnaseq " –metadata "platform,description,srx"
```

The `metahq retrieve` command uses ontological reasoning when executing queries, which enables users to query any tissue or disease term and retrieve not only samples directly annotated to that term, but also those annotated to all its descendant terms in the ontology (e.g., querying "heart" retrieves samples annotated to descendant terms such as "myocardial cell").

ML researchers can automatically convert queried tissue and disease annotations to ML-ready labeled sets, which goes beyond annotations that capture a sample or study's context to declare what a sample or study is, what it definitely is not, and what is uncertain. Such labeled sets can be used out-of-the-box for ML training and validation such as fine tuning and benchmarking large language model performance for various biomedical tasks.

### Term search

To ensure unambiguous annotation searches, users must query (with `metahq retrieve`) using ontology term IDs. In case the user does not know the ID for a term of interest, they can input a free-text query to find the IDs of the most relevant terms in a particular ontology. The following command will return the top five similar MONDO ontology terms when queried with "heart attack".

```
metahq search --query "heart attack" --type disease --ontology MONDO -k 5
```

## Materials and methods

### Harmonization

To harmonize the different sources, all annotations were mapped to the following standards: UBERON and CL ontology for tissues (and cell types), MONDO for diseases, 'male' and 'female' for sex, and seven unique age groups as defined in [15, 24] for age.

To minimize storage requirements, we reduced tissue and disease annotations to their most specific annotations. For example, if a sample was annotated to both "heart disorder" and "cardiomyopathy", then we dropped "heart disorder" since this annotation can be inferred through propagation with the ontology. Additionally, we dropped any samples or studies that only had annotations to general system-level terms (e.g., "cardiovascular system" or "immune system disorder"; see Supplemental File 1).

We also assigned evidence codes to all annotations in the database. Annotations were assigned evidence codes of either "expert" or "crowd" depending on whether the annotation was assigned by a domain expert or through crowdsourcing annotations by novices, respectively. Currently, the only annotations with "crowd" evidence codes are from CREEDS.

Metadata for all samples and studies were downloaded directly from GEO with FTP and the GEOfetch Python package [25]. Missing metadata were stored in the MetaHQ database as "NA". Mappings from GEO to SRA were obtained from the NCBI FTP server at: https://ftp.ncbi.nlm.nih.gov/sra/reports/ Metadata/SRA_Accessions.tab.

### Study-level annotations

Gemma is the only source that natively provided expert-curated study-level annotations. For all other sources, which provide only sample-level annotations, we derived study-level annotations using a simple inheritance rule: if any sample in a study had a particular annotation, we assigned that annotation to the entire study.

### Term search

We applied the fast BM25+ algorithm from the Rank-BM25 Python package [26] to calculate the similarity between a user's free-text query and all the terms in an ontology (based on term names and synonyms). BM25+ prioritizes exact matches to term names, followed by similarity to synonyms, weighted by the synonym's relevance to the term (e.g., exact or related).

### Propagating annotations and assigning labels

For the `metahq retrieve` command, when `--mode` is `annotate`, all samples (or studies) annotated to the query term or any of its descendants in the ontology are retrieved. When the mode is `label`, samples (or studies) directly annotated to the query term or any of its descendants are assigned a "+1" label (positive), those directly annotated to any of the query's ancestors are assigned a "0" label (neutral/unknown), and all other samples are assigned a "-1" label (negative). For example, if a query is "cardiomyopathy" used along with the `--mode` argument and passing `label` (`annotate` is the default), MetaHQ will provide a set of labels where samples directly annotated to "cardiomyopathy" or any of its descendants (such as "idiopathic cardiomyopathy" and "autoimmune cardiomyopathy") receive a positive label, and samples annotated to ancestor terms such as "heart disorder" receive a neutral label (because it is unclear which specific heart disorder those samples correspond to). Finally, samples annotated to all other terms, corresponding to those completely unrelated to the query (e.g., "liver cancer") will receive a negative label. The NetworkX Python package [27] was used to construct and operate on the ontology graphs.

MetaHQ also contains samples annotated as "control", which are incorporated into disease labels with a distinct label value. For example, if a user queries MetaHQ for "MONDO:0004994" (the term ID for "cardiomyopathy") with `--mode=label`, the samples annotated as "control" in the cardiomyopathy studies will receive a label of "2" for cardiomyopathy. We assign a value of 2 (rather than -1) to explicitly delineate which samples are verified controls versus samples that are negative because of their annotations to other diseases.

## Discussion

Currently, MetaHQ is limited to annotations for transcriptomics samples and studies in GEO and SRA. In future iterations, we aim to leverage its general software design to expand the database to include annotations to data from other omics types (e.g., epigenomics, proteomics, metabolomics, and metagenomics) and increase the number of types of annotations (e.g., phenotype, experimental factor) and their supported biomedical ontologies. To enable MetaHQ to grow beyond our collation efforts, we provide detailed instructions for researchers on how to contribute their own high-quality annotations to MetaHQ (https://meta-hq.readthedocs.io/en/latest/contribute/) We particularly encourage researchers to contribute sex and age annotations for samples that already have tissue and disease annotations in MetaHQ, though we note that these attributes are often unavailable in the original sample-level metadata and may not be possible to annotate [4].

Lastly, to recognize the painstaking efforts of curators who annotated all entries in MetaHQ, all samples and studies contain cross-references to their original sources (Table S1). We strongly encourage researchers using MetaHQ annotations to cite these original sources along with MetaHQ.

## Author contributions statement

AK, LV, and PH conceived and designed the project. PH, LV, KJ, CM, and SS curated the data, and PH and LV harmonized them. PH, FA, and CM wrote the software. PH wrote the manuscript, which was then reviewed by AK, CM, LV, FA, KJ, and SS.

## Acknowledgments

The authors thank Dave Bunten for reviewing the code and members of the Krishnan Lab for testing the software. This work is supported in part by funds from the National Science Foundation NSF BIO 2328140 to AK.

# Supplemental Figures

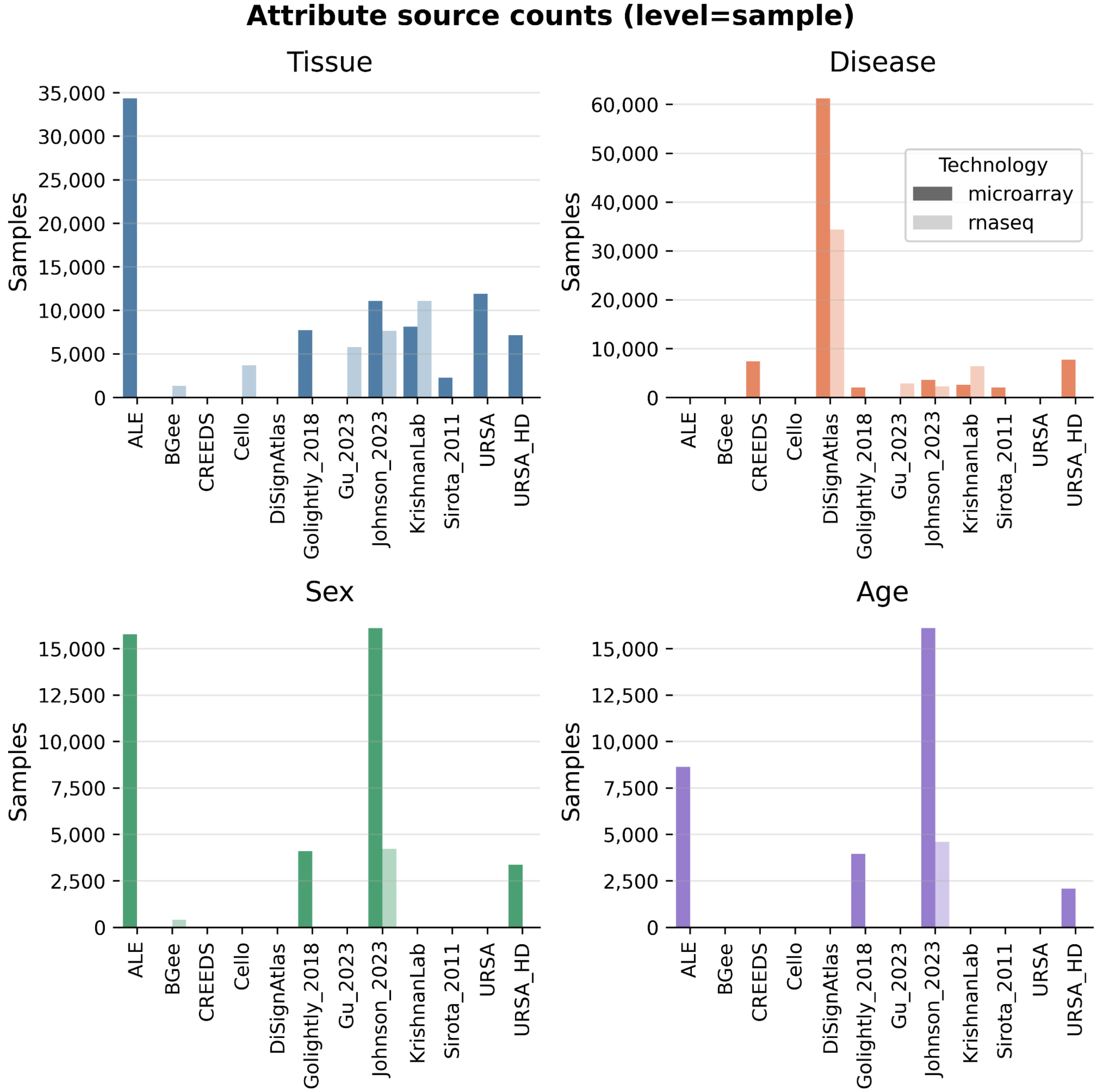

**Supplemental Figure S1**. The number of contributions made by each annotation source across attributes for sample-level annotations for RNA-Seq (light) and microarray (dark) samples.

**Supplemental Figure S2**. The number of contributions made by each annotation source across attributes for study-level annotations for RNA-Seq (light) and microarray (dark) samples.

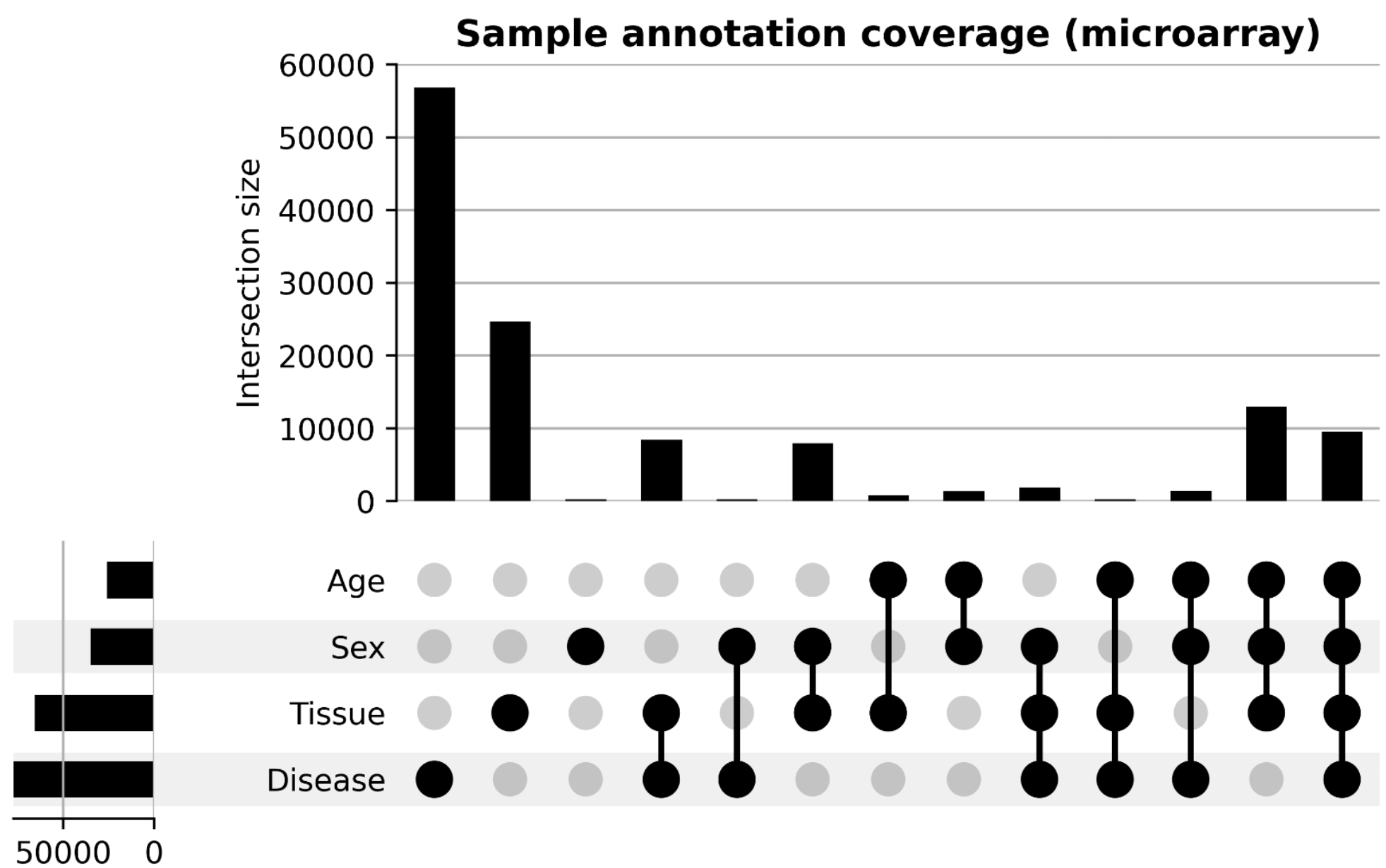

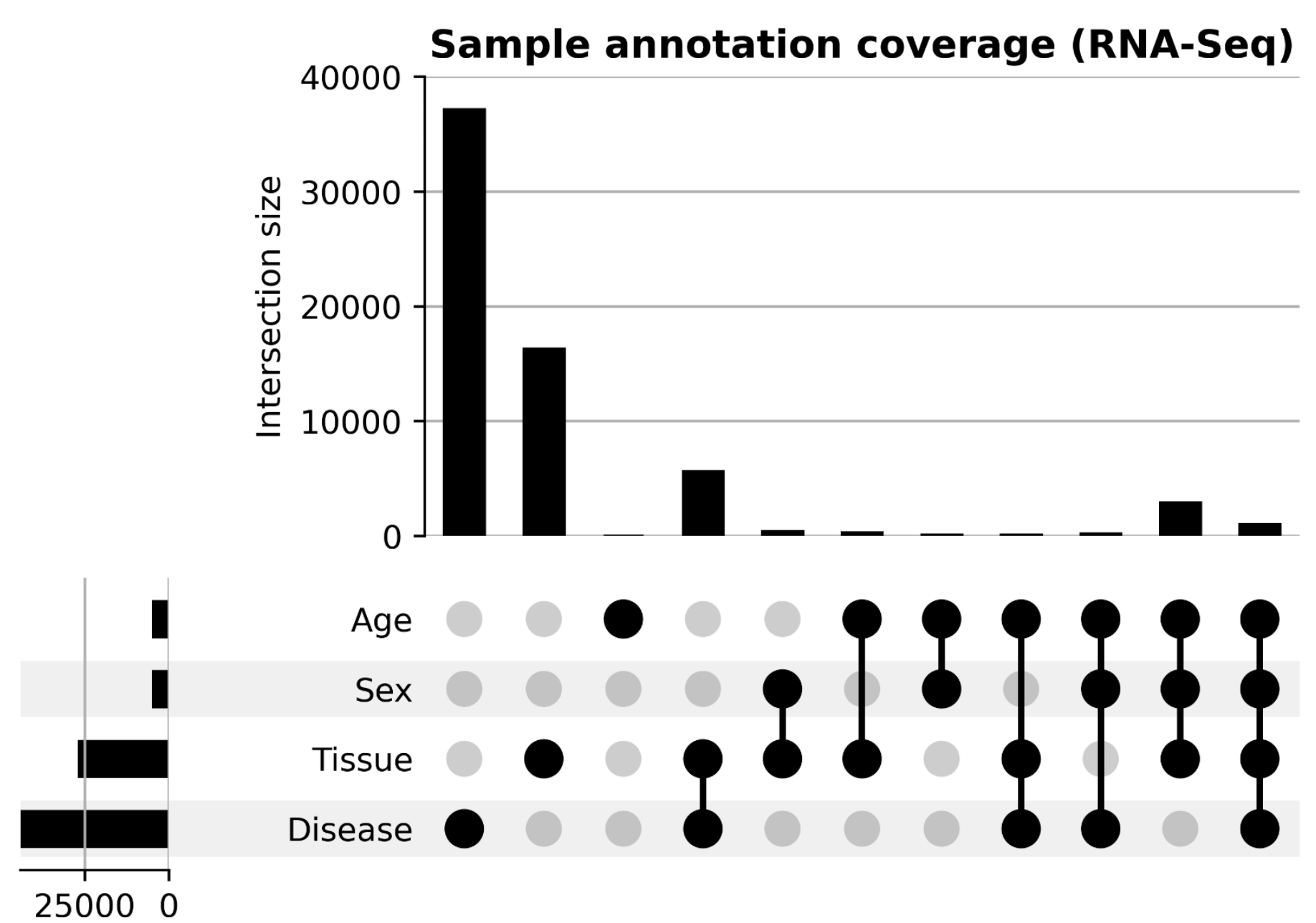

**Supplemental Figure S3**. Upset plot showing the coverage of attribute annotations per sample in MetaHQ for microarray (top) and RNA-Seq (bottom).

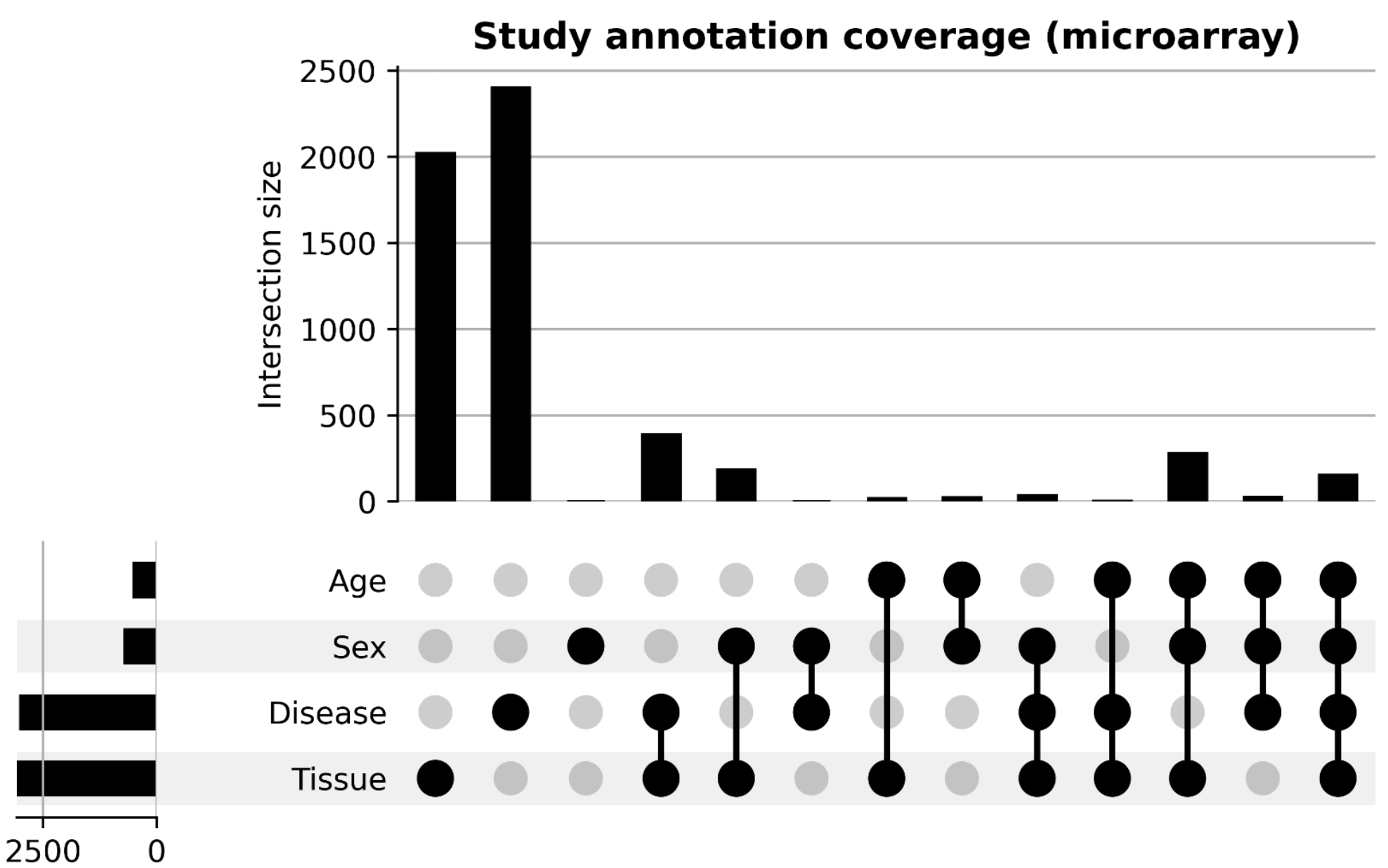

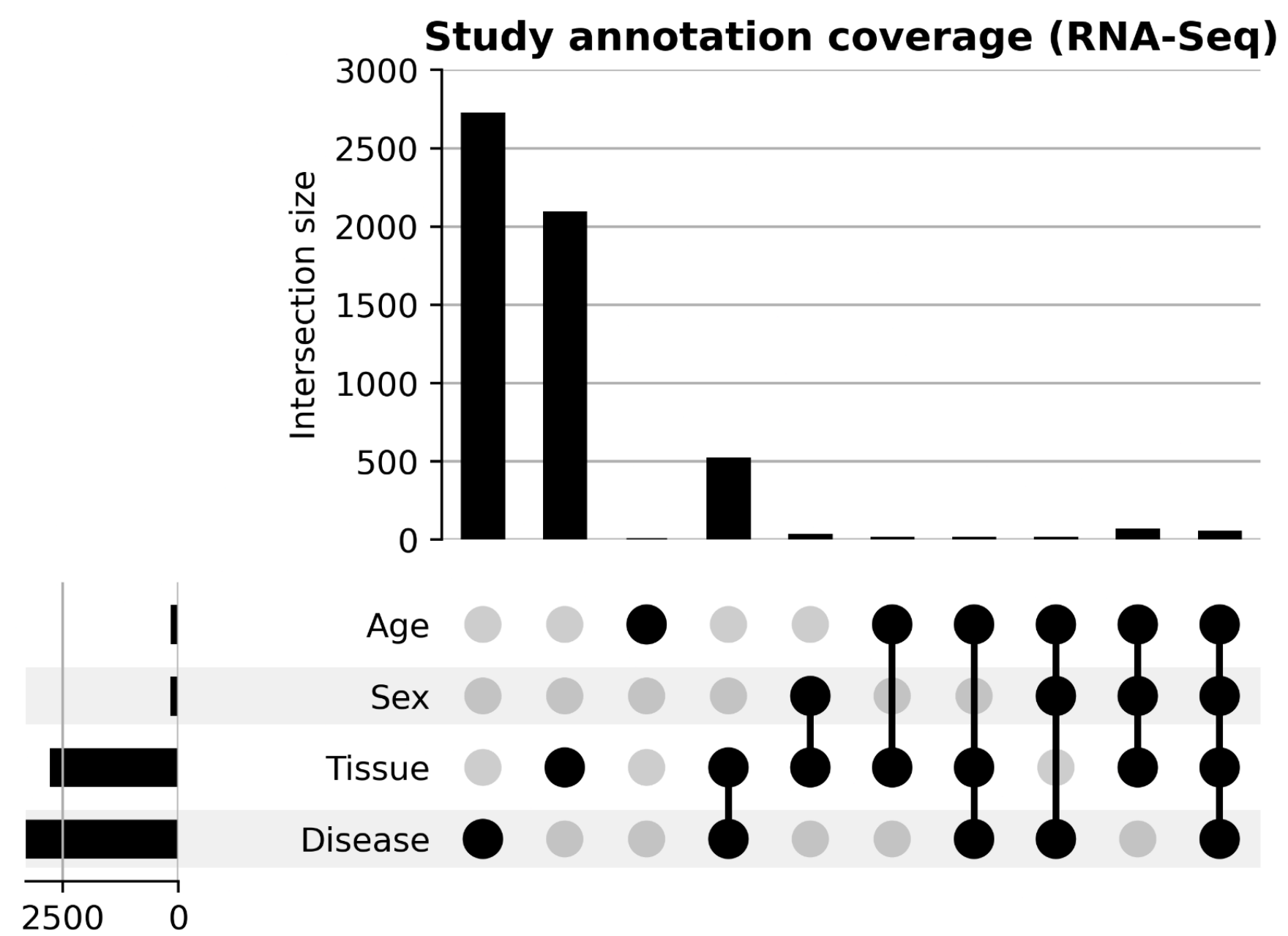

**Supplemental Figure S4**. Upset plot showing the coverage of attribute annotations per study in MetaHQ for microarray (top) and RNA-Seq (bottom).

**Supplemental Table S1**. The annotation sources harmonized in MetaHQ and their corresponding papers.

| Source | DOI |
| --- | --- |
| ALE | 10.1186/s12859-017-1888-1 |
| BGEE | 10.1093/nar/gkaa793 |
| CellO | 10.1016/j.isci.2020.101913 |
| CREEDS | 10.1038/ncomms12846 |
| DiSignAtlas | 10.1093/nar/gkad961 |
| Gemma | 10.1093/database/baab006 |
| Golightly_2018 | 10.1038/sdata.2018.66 |
| Gu_2023 | 10.1016/j.gpb.2021.08.017 |
| Johnson_2023 | 10.1101/2023.01.12.523796 |
| Sirota_2011 | 10.1126/scitranslmed.3001318 |
| URSA | 10.1093/bioinformatics/btt529 |
| URSA_HD | 10.1016/j.cels.2018.12.010 |

| tissues | diseases |
| --- | --- |
| UBERON:0000990 | MONDO:0020683 |
| UBERON:0001008 | MONDO:0002409 |
| UBERON:0002330 | MONDO:0002657 |
| UBERON:0002416 | MONDO:0045024 |
| UBERON:0002390 | MONDO:0004995 |
| UBERON:0004535 | MONDO:0019040 |
| UBERON:0001007 | MONDO:0003900 |
| UBERON:0002405 | MONDO:0004335 |
| UBERON:0001004 | MONDO:0700220 |
| UBERON:0001016 | MONDO:0021147 |
| UBERON:0000949 | MONDO:0002022 |
| UBERON:0002204 | MONDO:0024458 |
| UBERON:0000479 | MONDO:0005151 |
| UBERON:0000062 | MONDO:0005570 |
| UBERON:0001033 | MONDO:0003847 |
| UBERON:0004100 | MONDO:0043543 |
| UBERON:0005282 | MONDO:0700007 |
| UBERON:0005281 | MONDO:0005046 |
| UBERON:0008251 | MONDO:0005550 |
| UBERON:0003956 | MONDO:0021166 |
| UBERON:0005409 | MONDO:0002051 |
| UBERON:0001032 | MONDO:0005066 |
| UBERON:0004122 | MONDO:0044970 |
| UBERON:0002193 | MONDO:0006858 |
| UBERON:8450002 | MONDO:0002081 |
| UBERON:0001009 | MONDO:0005071 |
| UBERON:0001750 | MONDO:0005137 |
| UBERON:8600018 | MONDO:0700003 |
| UBERON:0007798 | MONDO:0100366 |
| CL:0000255 | MONDO:0024623 |
| CL:0000225 | MONDO:0100086 |
| CL:0000183 | MONDO:0029000 |
| CL:0000393 | MONDO:0021669 |
| CL:0000404 | MONDO:0019303 |
| CL:0002321 | MONDO:0002025 |
| CL:0000424 | MONDO:0043459 |
| CL:0000039 | MONDO:0005039 |
| CL:0000413 | MONDO:0005087 |
| CL:0000988 | MONDO:0002254 |
| CL:0000219 | MONDO:0043839 |
| CL:0002242 | MONDO:0044991 |
| CL:0000329 | MONDO:0002118 |
| CL:4033054 | |
| CL:0000412 | |
| CL:0011115 | |
| CL:0000151 | |
| CL:0007001 | |
| CL:0000325 | |
| CL:0000630 | |